\documentclass{nature}

\usepackage[top=0.8in, bottom=1in, left=0.8in, right=0.8in]{geometry}

\usepackage{graphicx}
\usepackage{amsmath}
\usepackage{amssymb}
\usepackage{threeparttable}
\usepackage{caption}
\usepackage{color}
\usepackage{adjustbox}
\usepackage{url}
\usepackage{soul}
\usepackage{utfsym} 
\usepackage{xcolor}
\usepackage{hyperref} 
\hypersetup{backref=true,       
    pagebackref=true,               
    hyperindex=true,                
    colorlinks=true,                
    breaklinks=true,                
    urlcolor= black,                
    linkcolor= blue,                
    bookmarks=true,                 
    bookmarksopen=false,
    filecolor=black,
    citecolor=blue,
    linkbordercolor=blue
}
\usepackage{lineno}

\newcommand{\um}{$\mu$m}

\title{A luminous and young galaxy at $z=12.33$ revealed by a JWST/MIRI detection of H$\alpha$ and [OIII]}
\author{}

\begin{document}
\maketitle

\vspace{-12pt}
\noindent\author{Jorge A. Zavala$^1$},
\author{Marco Castellano$^2$},
\author{Hollis B. Akins$^3$},
\author{Tom J. L. C. Bakx$^4$}, 
\author{Denis Burgarella$^5$},
\author{Caitlin M. Casey$^3$},
\author{{\'O}scar A. {Ch{\'a}vez Ortiz}$^3$}, \author{Mark Dickinson$^6$},
\author{Steven L. Finkelstein$^3$},
\author{Ikki Mitsuhashi$^{1,7}$},
\author{Kimihiko Nakajima$^1$},
\author{Pablo G. {P{\'e}rez-Gonz{\'a}lez}$^{8}$},
\author{Pablo Arrabal Haro$^6$},
\author{Pietro Bergamini$^{9,10}$},
\author{Veronique Buat$^5$},
\author{Bren Backhaus$^{11}$},
\author{Antonello {Calabr{\`o}}$^2$},
\author{Nikko J. Cleri$^{12,13}$},
\author{David {Fern{\'a}ndez-Arenas}$^{14,15}$}, 
\author{Adriano Fontana$^2$}, 
\author{Maximilien Franco$^3$},
\author{Claudio Grillo$^{9,16}$},
\author{Mauro Giavalisco$^{17}$},
\author{Norman A. Grogin$^{18}$}, 
\author{Nimish Hathi$^{18}$},
\author{Michaela Hirschmann$^{19,20}$},
\author{Ryota Ikeda$^{1,21}$},
\author{Intae Jung$^{18}$},   
\author{Jeyhan S. Kartaltepe$^{22}$},  
\author{Anton M. Koekemoer$^{18}$},   
\author{Rebeca L. Larson$^{3,22}$},  
\author{Jed McKinney$^{3}$},
\author{Casey Papovich$^{12,13}$},
\author{Piero Rosati$^{23, 10}$}, 
\author{Toshiki Saito$^{1}$}, 
\author{Paola Santini$^2$},
\author{Roberto Terlevich$^{24,25,26}$},
\author{Elena Terlevich$^{24,26}$}, 
\author{Tommaso Treu$^{27}$}, and
\author{L. Y. Aaron {Yung}$^{18}$}\\
\vspace{-10pt}
\noindent\author{(\it Affiliations are included at the end of the document)}\\





\vspace{10pt}
\begin{abstract}
The James Webb Space Telescope (JWST) has discovered a surprising population of bright galaxies in the very early universe ($\lesssim$ 500 Myrs after the Big Bang) that is hard to explain with conventional galaxy formation models and whose physical properties remain to be fully understood. Insight into their internal physics is best captured through nebular lines but, at these early epochs, the brightest of these spectral features are redshifted into the mid-infrared and remain elusive.
Using  the JWST Mid-Infrared Instrument, MIRI, here we present the first detection of H$\mathbf{\alpha}$ and doubly-ionized oxygen ([OIII]4959,5007\,\AA) at $\mathbf{z>10}$.  These detections place the bright galaxy GHZ2/GLASS-z12 at $\mathbf{z=12.33\pm0.04}$, making it the most distant astronomical object with direct spectroscopic detection of these lines. These observations provide key insights into the conditions of this primeval, luminous galaxy, which shows hard ionizing conditions rarely seen in the local Universe likely driven by compact and young ($\mathbf{\lesssim30\,}$Myr) burst of star formation. 
Its oxygen-to-hydrogen abundance is close to a tenth of the solar value, indicating a rapid metal enrichment. 
This study confirms the unique conditions of this remarkably bright and distant galaxy and the huge potential of mid-IR observations to characterize these objects.
\end{abstract}

Following the confirmation of the first surprisingly bright galaxies at high-redshifts ($z\gtrsim10$)\cite{ArrabalHaro2023,Bunker2023,Harikane2024}, research efforts must now shift towards gaining a deeper understanding of their physical properties. 
The detection and interpretation of emission-line spectra are pivotal in this topic, with well-studied calibrations and diagnostic diagrams based on rest-frame optical transitions and line ratios such as [NII]/H$\alpha$ vs [OIII]/H$\beta$. 
The advent of the JWST and, particularly, the sensitive Near Infrared Spectrograph (NIRSpec), has now unlocked access to some of these lines in very high-redshift galaxies, enabling detailed studies that were previously unreachable beyond $z\sim3$ (e.g.\cite{Trump2023,Sanders2023}). 
At redshifts above $z\sim7$, however, the H$\alpha$ transition, known as the gold standard to measure young star formation activity, is redshifted beyond the NIRSpec coverage. Similarly,  the [OIII] and H$\beta$ lines, sensitive to metallicity and ionizing conditions,  can only be observed up to $z\sim9.5$ with this instrument. 

The MIRI instrument on board JWST is thus the only astronomical instrument with the required wavelength coverage to detect these spectral lines at higher redshifts, critical for characterizing the physical properties of the first galaxies in the universe. While early predictions suggested that their successful detection may require long observing times of several tens of hours\cite{Rieke2015}, the combination of both the better-than-expected performance of JWST\cite{Rigby2023} and the remarkably high brightness of some od the JWST-discovered distant galaxies, might have improved the prospects for such studies, making the detection of $z>10$ rest-frame optical emission lines more feasible than previously anticipated. 

Here, we report the results of the first MIRI spectroscopic observations on a $z>10$ galaxy candidate and test its efficiency for redshift confirmation and characterization of early galaxies via the detection of rest-frame optical nebular lines.

\begin{figure*}[h]
  \begin{center}
  \includegraphics[width=\linewidth]{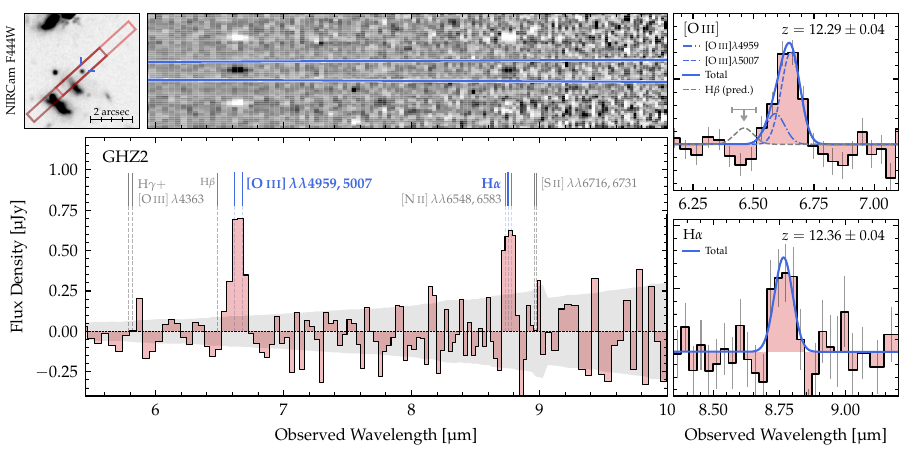}
  \end{center}
    \vspace{-0.6truecm}  
    \caption{\small \textbf{JWST/MIRI spectrum of GHZ2 at $z=12.33$.} {\bf Top left:} NIRCam F444W cutout image ($5''\times5''$) centered at the position of GHZ2, with the MIRI/LRS slit illustrated with the red rectangle (at the two different dither positions). The combined 2D spectrum and the used aperture for the 1D extraction are also plotted (see details in the Section Methods).  {\bf Bottom left:} 1D extracted spectrum at the position of GHZ2 across the most sensitive wavelength range, $\lambda_{\rm obs}\approx5.7-10\,\micron$ and the associated 1$\sigma$ uncertainty (gray region). The expected wavelengths of typically bright rest-frame optical emission lines are indicated with the vertical lines, with the blue text highlighting the robust detections. {\bf Right:} 
     Zoom-in on the detected spectral features identified as the [OIII]4959,5007\AA\ doublet ({\it top})  and the H$\alpha$ emission line ({\it bottom}) along with the best-fit Gaussian functions and the implied redshifts and associated $1\sigma$ uncertainties, which are dominated by the current wavelength calibration.  In the top panel, the 2.5$\sigma$ upper limit for the H$\beta$ line is indicated with the 
     downward arrow, while the predicted H$\beta$ line from the Balmer decrement (assuming a dust-free scenario) is represented by the 
      gray dashed line. }
   \label{fig:MIRI_spectrum}
\end{figure*}

We target the galaxy GHZ2/GLASS-z12\cite{Castellano2022,Naidu2022} ($\rm RA=00^h13^m59.76^s$, $\rm DEC=-30^o19'29.11\arcsec$) with MIRI to search for the brightest rest-frame optical nebular emission lines redshifted into the mid-IR regime: [OIII]4959,5007\AA, $H\alpha$, $H\beta$ (with the doublets [NII]6548,6583\AA\ and [SII]6716,6731\AA\ also covered but expected to be too faint to be detected). 
The target stands out as one of the brightest and most robust among the extremely distant galaxy candidates found in deep JWST NIRCam images, with a photometric redshift of $z\approx12.0-12.4$. It was found in the outskirts of the galaxy cluster Abell 2744, as part of the GLASS-JWST Early Release Science Program\cite{Treu2022a}, with an absolute rest-frame UV magnitude of $M_{\rm UV} = -20.5$~mag\cite{Castellano2024} and a inferred stellar mass\cite{Naidu2022} close to $10^9\,M_\odot$, despite its  compact morphology\cite{Ono2023} ($R_e\lesssim50\,$pc) and  moderate gravitational amplification of\cite{Bergamini2023} $\mu\approx1.3$ (see Methods for a deeper discussion on the gravitational amplification).

The MIRI observations were conducted on October 25-29, 2023, using the low resolution spectrometer (LRS) slit mode (resolving power $R\approx50-200$), with a total integration on-source exposure time of 9\,h (see details in the Methods). The main advantage of this mode relies on its large spectral coverage, which provides sensitive observations across $\approx5–12\,\mu$m, covering at least one of the aforementioned lines across $z\approx7-20$. Data reduction was performed using the standard JWST pipeline with some additional steps as described in the Methods.

Fig. \ref{fig:MIRI_spectrum} shows the   MIRI one-dimensional (1D) and two-dimensional (2D) spectra of the source. Two spectroscopic features are clearly detected above the noise level ($>5\sigma$) in the 1D and 2D spectra. These spectral features are associated with the [OIII] doublet (4959\AA\ and 5007\AA) and with the H$\alpha$ recombination line, which constrain the redshift of this source to be $z_{\rm spec}=12.33\pm0.04$ (see also ref.\cite{Castellano2024}), making this object the most distant galaxy with detections of these nebular lines and one of the brightest early galaxies discovered to date.
The JWST/NIRSpec spectra of the other five galaxies with similar spectroscopic redshifts\cite{Curtis-Lake2023,DEugenio2023,Wang2023,Carniani2024} ($z\approx12-14$) are distinguished by the lack of strong emission lines. In these cases, the spectroscopic redshifts are measured via the spectral break in continuum emission produced by absorption of neutral hydrogen, but extensive tests are needed to rule out lower-redshift solutions that can produce similar breaks.
This demonstrates the unique use of MIRI to spectroscopically confirm the highest redshift galaxies via direct detection of rest-frame optical lines that can additionally provide direct information on the galaxy's star formation rate and ionized gas properties, although these studies might be limited to the brightest systems.

The emission line around $6.6\,\micron$, associated with the [OIII] doublet, was fitted with a double Gaussian function to infer the line flux densities (reported in Table 1). During this procedure we assume the same line-width for the two lines and fix the relative intensity ratio to the theoretical 1:3 value\cite{Storey2000}. In the case of  H$\alpha$, the line was fitted with a single Gaussian function. Although the [NII] doublet lines are blended with H$\alpha$, at the redshift of GHZ2,  it is reasonable to assume that the flux contribution of [NII] is negligible due to the likely sub-solar metallicity\cite{Groves2006,Nakajima2022b}. Negligible [NII] emission is seen even at lower redshifts ($z\approx4-7$) for galaxies\cite{Cameron2023,Sanders2023,Calabro2024,Topping2024} and AGN alike\cite{Kocevski2023,Harikane2023},  an even for for nitrogen-enriched systems\cite{Topping2024}. 
Hence, we can safely assume that this line is dominated by the H$\alpha$ emission.
For the undetected emission lines, including H$\beta$ and the [SII] doublet, $2.5\sigma$ upper limits were derived by adopting the local noise r.m.s and a given line-width as described in the Methods. Note that the non-detection of H$\beta$ is still consistent with theoretical predictions and with zero dust attenuation. A more stringent upper limit on the H$\beta$ flux density can be inferred from the measured H$\alpha$ flux density, adopting a flux ratio of $\rm H\alpha/H\beta=2.85$ (based on the so-called case-B recombination scenario and under the typical physical conditions of galaxies' ionized gas\cite{Osterbrock2006}).
This ratio is valid in the case of negligible dust attenuation, but it increases in the presence of dust since the H$\beta$ line is more affected by dust extinction. Nevertheless, the dust attenuation is not expected to be significant in this galaxy (with inferred $\rm A_V$ values around or below 0.1-0.3\,mag, as described in the Methods).

\begin{figure*}[h]
  \begin{center}
  \includegraphics[width=\linewidth]{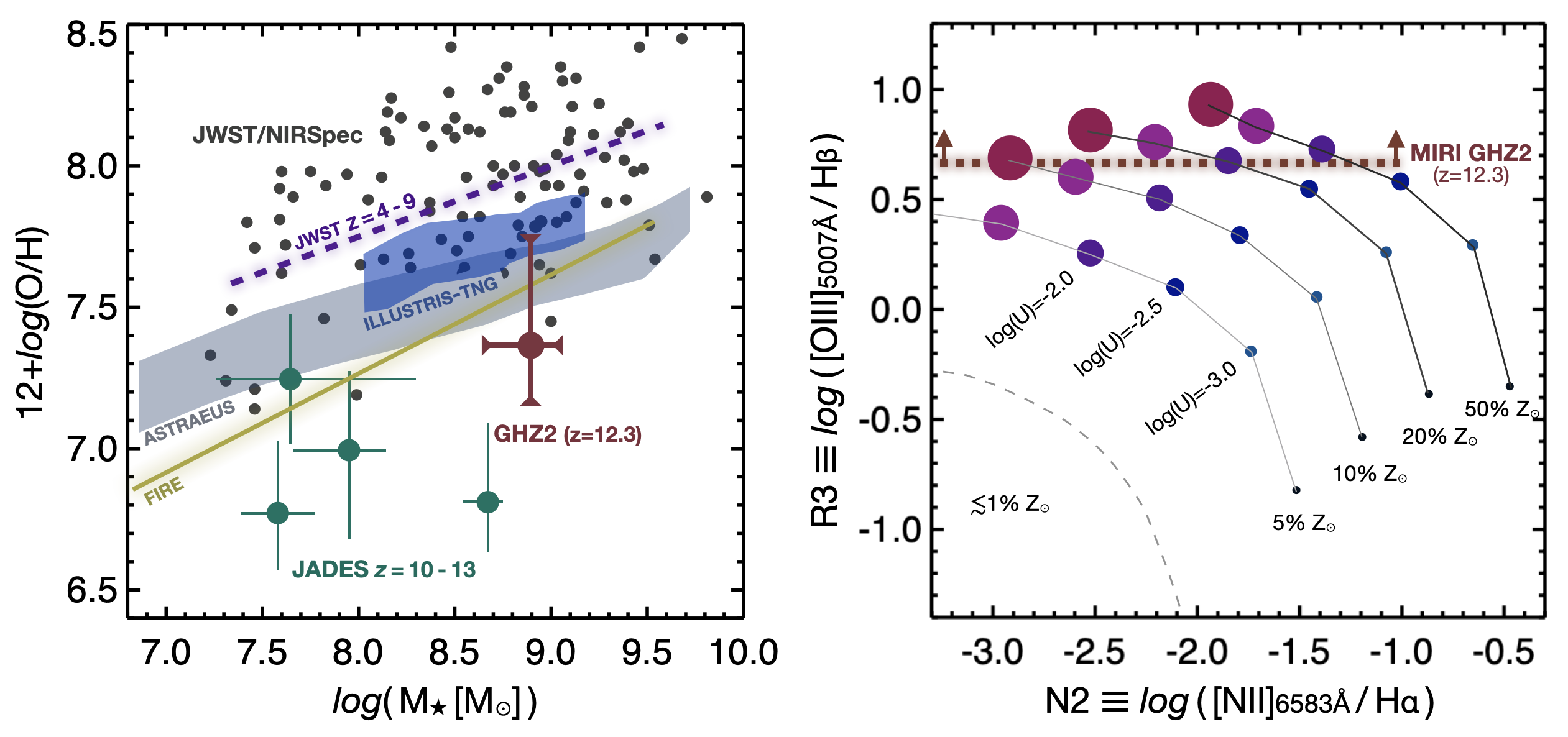}
  \end{center}
    \vspace{-0.6truecm}  
    \caption{\small \textbf{The stellar mass-metallicity relation and the ionization diagnostic diagram.} {\bf Left:} Position of GHZ2 in the stellar mass-metallicity plane in comparison to lower redshift galaxies at  $z\approx4-9$ (black dots for individual galaxies and purple line for best-fit relationship\cite{Nakajima2023}) and with the $z=10-13$ galaxies from the JADES survey\cite{Curtis-Lake2023} (green points; metallicity estimated from SED fitting). Error bars represent $1\sigma$ uncertainties in the stellar mass values and the range of allowed values for the metallicities. All the $z>10$ galaxies, including GHZ2, are off from the lower-redshift relationship\cite{Nakajima2023}, suggesting an evolution towards lower metallicity values at earlier epochs.  The predictions from the FIRE\cite{Ma2016} (golden line) and  ASTRAEUS\cite{Ucci2023} (gray region) simulations at $z\sim10$ show a broad agreement with the current constraints, while the ILLUSTRIS-TNG\cite{Torrey2019} simulations (blue region) predict slightly higher metallicities, although still consistent within the error bars. Despite the early epochs, none of these galaxies show pristine (close-to-zero) metallicities, implying a rapid metal enrichment in the early Universe.
    {\bf Right:} The $\rm [NII]/H\alpha$ vs $\rm [OIII]/H\beta$  diagram along with the predictions from a stellar photoionization model\cite{Nakajima2022b} assuming four different metallicities and six ionization parameters in steps of 0.5dex (models with $Z\lesssim0.01\,Z_\odot$ would lie below the dashed line). The observational constraints on the R3 value (formally a lower limit derived under the assumption of zero dust attenuation) implies  high ionizing conditions, with $\log(U)>-2.0$. 
    }
   \label{fig:MZR}
\end{figure*}

This record-breaking detection of H$\alpha$ provides a direct probe of the young star formation activity, tracing massive stars with ages around or below $10\,$Myr. This, combined with the detection of oxygen (revealing the presence of metals and thus of more evolved stars), provides a unique opportunity to study the stellar population of this distant galaxy and its ionizing photon production efficiency. We infer the average stellar age, and other properties like stellar mass and star formation rate, by conducting spectral energy distribution (SED) fitting to the NIRCam photometry jointly with the constraints from the H$\alpha$ and [OIII] emission lines (see details in the Methods and Supplementary materials). 
The photometry and the spectroscopic data are well reproduced with a model with a composite star formation history extending for $\sim$50~Myr, with a mass-weighted age of $28^{+10}_{-14}\,$Myr, and more than 60\% of the total stellar mass formed during the past $30\,$Myr.
The presence of young and massive stars implies a higher rate of ionizing photon production compared to typical values in galaxies at lower redshifts (see Figure \ref{fig:ion_efficiency}).
The evidence that most of the stellar mass of GHZ2 formed recently
is consistent with other results implying that early galaxies have more ``bursty" star-formation histories\cite{endsley23,ciesla23,tacchella22,cole2023}).  If the majority of early galaxies do indeed form the bulk of their visible mass in their recent past, it could explain not only the remarkable luminosity of this distant galaxy, but the overall surprising number of observed bright galaxies in this epoch\cite{finkelstein23,adams23}.

The SED-based SFR (averaged on the last 10\,Myr and taking into account the gravitational lensing amplification of\cite{Bergamini2023} $\mu=1.3$) of $5\pm2\rm\,M_\odot\,yr^{-1}$ is in good agreement with the SFR of $9\pm3\rm\,M_\odot\,yr^{-1}$ estimated directly from the H$\alpha$ luminosity assuming the calibration from ref.\cite{Reddy2022}. This calibration is based on low-metallicity stellar population synthesis models that include the effects of massive stars in binary systems characterized by a high ionizing photon production efficiency. On the other hand, the widely-adopted calibration used for local and low-redshift galaxies\cite{Kennicutt2012} predicts a higher SFR of $\sim22\pm5\,\rm M\odot\,yr^{-1}$, mainly due to the absence of these low-metallicity and binary stars. 

The JWST/MIRI data also constrain  the R3~$\equiv \log(\rm[OIII]/H\beta)$ ratio, which is known to correlate with the gas-phase metallicity\cite{Nagao2006}. 
The estimated line ratio and its associated uncertainty of $\rm [OIII]/H\beta=5.2\pm 1.5$ (when assuming directly the inferred H$\beta$ in the case of zero dust attenuation) implies a metallicity of $12+\log\rm(O/H)=7.40^{+0.52}_{-0.37}$ according to the relation presented by ref.\cite{Sanders2024}, corresponding to $Z=0.05^{+0.12}_{-0.03}\,Z_\odot$ (where $Z_\odot$ is solar metallicity).  A similar range of metallicities are obtained when using the theoretical calibrations\cite{Hirschmann2023} specifically designed for galaxies at $z>4$ (see details in the Methods).  
These values are in  good agreement with the independent metallicity estimation of  $12+\log\rm(O/H)=7.26^{+0.27}_{-0.24}$ based on the [NeIII]3868\AA/[OII]3727\AA\ index (with additional constraints from other lines), as reported in our parallel analysis of the NIRSpec data of this source\cite{Castellano2024}.

Despite the young age derived for GHZ2, it is notable that its metallicity is already enriched to a few percent (or even up to $\sim10-15\%$) of the solar value, significantly above expectations for the primordial objects dominated by the first-generation stars (typically known as  population-III stars). This implies a very rapid metal enrichment during the earliest phases of galaxy formation. The metallicity inferred for GHZ2 is higher than the metallicities of the four spectroscopically-confirmed galaxies at $z=10-13$ discovered in the JADES survey\cite{Curtis-Lake2023}. This is not totally surprising since a correlation between stellar mass and metallicity is known to exist even up to $z\sim9$\cite{Nakajima2023}, and the stellar mass of GHZ2 of $\log(M_\star/M_\odot)\approx8.6-9.0$ is around an order of magnitude larger than what was inferred for JADES galaxies. This might suggest that a similar relation exists even at these early redshifts, although shifted to lower metallicites. Actually, when compared with the so-called fundamental metallicity relation\cite{Curti2020}, which involves the SFR as a third parameter, all these $z>10$ galaxies deviate significantly towards lower metallicity values. Confirming the existence of such scaling relations would require, however, larger samples of galaxies with similar spectroscopy data.

\begin{figure}[h]
  \begin{center}
  \includegraphics[width=1.04\linewidth]{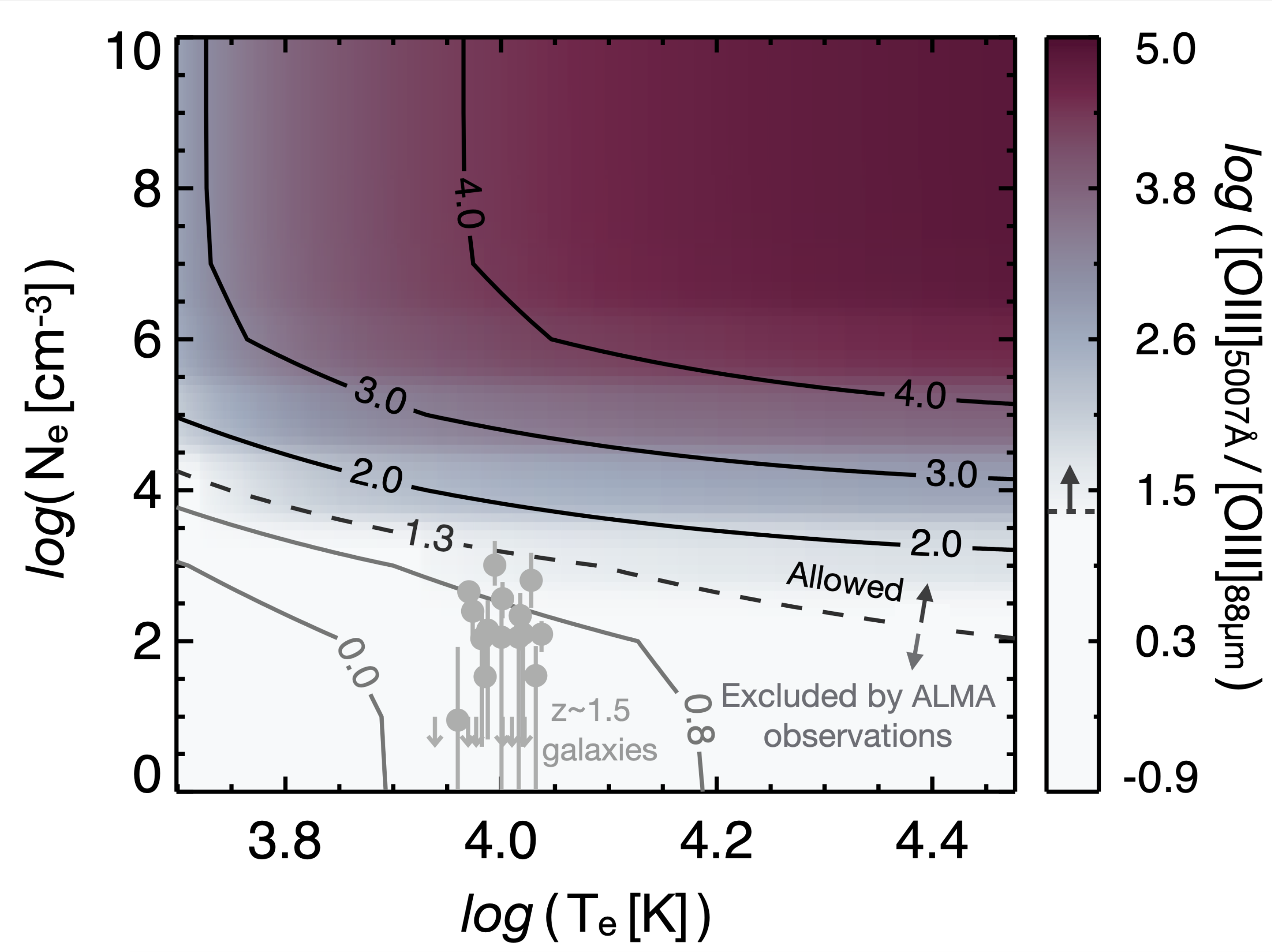}
  \end{center}
    \vspace{-0.6truecm}  
    \caption{\small \textbf{Electron density constraints.} Predicted [OIII]5007\AA-to-[OIII]88$\mu$m line ratio (illustrated in different colors and enclosed by the contours) as a function of electron density ($n_e$; y-axis) and electron temperature ($T_e$; x-axis). The constraints from the available ALMA observations in combination with the JWST/MIRI data, with $\log(\rm[OIII]_{5007A}/[OIII]_{88\mu m})\gtrsim1.3$, imply high electron densities above
    even above $1,000\,\rm cm^{-3}$ at the typically adopted electron temperature of $T_e=10,000\,$K (with a minimum electron density of $100\,\rm cm^{-3}$, depending on $T_e$). This is higher than the typical values measured at lower redshifts,  corroborating the extreme conditions of this early galaxy. For comparison, we plot the estimated electron densities (and associated $1\sigma$ uncertainties) for a sample of $z\sim1.5$ galaxies with direct $n_e$ measurements from the [OII] doublet\cite{Kaasinen2017} ($T_e$ were scattered around the adopted 10,000\,K for better visualization).
    }
   \label{fig:ALMA_constraints}
\end{figure}

To gain further insights into the physical conditions of the ionized gas in this galaxy, we show in Fig. \ref{fig:MZR} the well-studied $\rm [NII]/H\alpha$ vs $\rm [OIII]/H\beta$ BPT-diagram along with the predictions from a photoionization model powered by star formation\cite{Nakajima2022a}.  The R3 value of $5.2\pm 1.5$ derived above implies a high  ionization parameter of $\log(U)\gtrsim-2.0$, as shown in Fig. \ref{fig:MZR}. Note that this measurement is not affected by our assumption of a dust-free environment when calculating the H$\beta$ line since any dust attenuation will only increase the inferred [OIII]/H$\beta$ ratio.
The high ionization conditions in this galaxy are also confirmed by the detection of NIV]1488\AA\ and HeII 1640\AA in the NIRSpec spectrum\cite{Castellano2024}, requiring ionizing photons with energy above $\sim54\,$eV. This might point towards the presence of X-ray binaries or very massive stars\cite{Schaerer2019,Lecroq2024,Upadhyaya2024} contributing to the hard ionizing radiation.  

Alternatively, the high ionization conditions might be produced by AGN activity, although the higher ionization AGN tracers NeIV]2424\AA\ and  [NeV]3426\AA\ remain undetected\cite{Castellano2024}. On the other hand, a morphological analysis of this object constrains its size to be very compact ($r_e\lesssim50\,$pc) but marginally resolved\cite{Ono2023}, which favors the star-formation scenario as the source of the ionizing radiation or a composite AGN/star-forming galaxy system (rather than pure AGN).

It is now useful to put into context the upper limit on the [OIII]88$\mu$m line luminosity derived from previous ALMA observations\cite{Bakx2023}. The line ratio between this and the [OIII]5007\AA\ transition is sensitive to the electron density of the ionized gas with an extra milder dependency on temperature.
Fig. \ref{fig:ALMA_constraints} shows the expected [OIII]5007\AA-to-[OIII]88$\mu$m  line ratio as a function of electron temperature and density (see Methods section for further details). As can be seen, at the typical [OIII] electron temperatures (e.g. 6,000-15,000\,K), the constraints imposed by the ALMA and JWST observations imply a high electron density above $10^3\,\rm cm^{-3}$.   This, again, contrasts with the typical conditions seen in the local Universe and lower redshift galaxies with average densities on the order of\cite{Kaasinen2017} $n_e\approx10^1-10^2\,\rm cm^{-3}$, but  is in agreement with the recent results proving a redshift evolution  towards higher electron densities at high redshifts\cite{Isobe2023} (see also Figure \ref{fig:ne_redshift}).
This evolution might be driven by the
higher star formation rate surface density ($\Sigma_{\rm SFR}$) measured in high-redshift galaxies\cite{Reddy2023,Calabro2024,Topping2024}, which reaches an extreme value of $320\pm130\rm\,M_\odot\,yr^{-1}\,kpc^{-2}$ in GHZ2 (adopting the measured\cite{Ono2023} effective radius of 50\,pc). The combination of these unique conditions, including  high ionization parameters, high electron densities, and high star formation rate surface density might explain the brightness of the unique population of $z>10$ galaxies recently discovered by JWST, along with the young stellar ages and relatively low metallicities.

This study demonstrates the enormous potential of the Mid-Infrared Instrument, MIRI, on board JWST for the confirmation and characterization of the most distant galaxies in the Universe.
Our observations make our target, GHZ2, the most distant galaxy with direct detection of several transitions from the ionized gas (see also ref.\cite{Castellano2024}) and one of the brightest spectroscopically-confirmed galaxy at this early epoch, with a robust spectroscopic redshift of $z=12.33\pm0.04$.
The physical conditions of GHZ2, revealed directly by nebular emission lines, are extreme and rarely seen in the local Universe, with a low (but not pristine) metallicity, high ionization conditions, and high electron density. This emerging picture of compact galaxies with extreme conditions seems to be fairly common at high redshifts, particularly among the brightest systems, and might be associated with short bursts of  young star formation with ages of a few tens of Myr and with the presence of massive, low-metallicity stars. Some of the  properties of this early galaxy might also resemble the observed features of AGN, particularly the hard ionizing spectra inferred from the observations presented here and from the JWST/NIRSPec data\cite{Castellano2024}. It is thus possible that this source might be, at least partially, powered by an active black hole. Further observations on this and other similar sources will significantly contribute to  our understanding of early galaxy formation and black-hole growth, pushing the current frontiers into the formation epoch of the first massive objects in the Universe.\\

{\bf Data availability.} 
The JWST/MIRI data used in this paper will be publicly available through the Mikulski Archive for Space Telescopes server, under JWST program GO-3703. All other data generated throughout the analysis are available from the corresponding author, JAZ, on request.

{\bf Acknowledgements.} 
This work is based on observations made with the NASA/ESA/CSA James Webb Space Telescope. The data were obtained from the Mikulski Archive for Space Telescopes at the Space Telescope Science Institute, which is operated by the Association of Universities for Research in Astronomy, Inc., under NASA contract NAS 5-03127 for JWST. These observations are associated with program JWST-ERS-3703. This paper makes use of the following ALMA data:
ADS/JAO.ALMA\#2021.A.00020.S. ALMA is a partnership of ESO
(representing its member states), NSF (USA), and NINS (Japan),
together with NRC (Canada), MOST, ASIAA (Taiwan), and KASI
(Republic of Korea), in cooperation with the Republic of Chile. 
The Joint ALMA Observatory is operated by ESO, AUI/NRAO, and NAOJ. JAZ acknowledge funding from  JSPS KAKENHI grant number KG23K13150.
The GLASS-JWST team and other co-authors acknowledges support by NASA via grants JWST-ERS-1342 and JWST-GO-3703. C.G. acknowledges financial support through grant PRIN-MUR 2020SKSTHZ. Finally, we would like to thank Sarah Kendrew, Greg Sloan, and Milo Docher for their support during the preparation of the observations and their suggestions for data reduction.

{\bf Author Contributions.}
J.A.Z. led the JWST MIRI observing proposal, data analysis, and manuscript writing. M.C., A.C., A.F., P.S. and T.T. led the original JWST GLASS observations and the discovery study of GHz2, and contributed to the MIRI follow-up proposal. H.B.A and J.M. significantly contributed to the MIRI data reduction. T.J.L.C.B. and I.M. re-analyzed the ALMA data and the [OIII]88\micron constraints. D.B., O.A.C.O., S.L.F., P.P.G., V.B. and P.S. modelled and interpreted the spectral energy distributions of our target. P.B., C.G., and P.R. performed the lensing analysis presented in the Methods. M.D., C.M.C., P.A.H., M.F., M.G., N.A.G., N.H., R.I., I.J., J.S.K, A.M.K., R.L.L. and C.P., contributed to the original JWST MIRI proposal, observing design, and to the scientific discussion in the  proposal and manuscript. M.C., K.N., A.C., B.B., N.J.C., D.F.A., C.P., R.T., E.T. and T.S. focused on the interpretation of the observed lines and line ratios (and associated derived physical properties). K.N., M.H. and L.Y.A.Y. provided the theoretical grounds to infer some of the physical parameters and other predictions from simulations. All co-authors contributed to the editing and formatting of the manuscript.

{\bf Competing interests.}
The authors declare no competing interests.

\vspace{2pt}
\noindent\rule{\linewidth}{0.4pt}
\vspace{2pt}

\clearpage
\noindent{\bf \large Affiliations}\vspace{-20pt}
\begin{affiliations}
\small
 \item National Astronomical Observatory of Japan, 2-21-1, Osawa, Mitaka, Tokyo, Japan
 \item INAF - Osservatorio Astronomico di Roma, Via Frascati 33, 00078, Monte Porzio Catone, Italy 
 \item Department of Astronomy, The University of Texas at Austin, 2515 Speedway Boulevard Stop C1400, Austin, TX 78712, USA
 \item Department of Space, Earth, \& Environment, Chalmers University of Technology, Chalmersplatsen 4, SE-412 96 Gothenburg, Sweden
 \item Aix Marseille Univ, CNRS, CNES, LAM, Marseille, France
 \item NSF's National Optical-Infrared Astronomy Research Laboratory, 950 N. Cherry Avenue, Tucson, AZ 85719, USA
 \item Department of Astronomy, The University of Tokyo, 7-3-1 Hongo, Bunkyo, Tokyo 113-0033, Japan 
 \item Centro de Astrobiología (CAB), CSIC-INTA, Ctra. de Ajalvir km 4, Torrejón de Ardoz, E-28850, Madrid, Spain
 \item Dipartimento di Fisica, Universit\`a degli Studi di Milano, Via Celoria 16, I-20133 Milano, Italy
 \item INAF – OAS, Osservatorio di Astrofisica e Scienza dello Spazio di Bologna, via Gobetti 93/3, I-40129 Bologna, Italy
 \item Department of Physics, 196A Auditorium Road, Unit 3046, University of Connecticut, Storrs, CT 06269, USA
 \item Department of Physics and Astronomy, Texas A\&M University, College Station, TX 77843-4242 USA   
 \item George P. and Cynthia Woods Mitchell Institute for Fundamental Physics and Astronomy, Texas A\&M University, College Station, TX 77843-4242 USA 
 \item Canada-France-Hawaii Telescope, Kamuela, HI 96743, USA
 \item Instituto de Radioastronom{\'i}a y Astrof{\'i}sica, UNAM Campus Morelia, Apartado postal 3-72, 58090 Morelia, Michoac{\'a}n, Mexico
 \item INAF -- IASF Milano, via A. Corti 12, I-20133 Milano, Italy 
 \item University of Massachusetts Amherst, 710 North Pleasant Street, Amherst, MA, 01003-9305, USA 
 \item Space Telescope Science Institute, 3700 San Martin Dr, Baltimore, MD 21218, USA
 \item Institute for Physics, Laboratory for Galaxy Evolution and Spectral Modelling, Ecole Polytechnique Federale de Lausanne, Observatoire de Sauverny, Chemin Pegasi 51, 1290 Versoix, Switzerland
 \item INAF, Osservatorio Astronomico di Trieste, Via Tiepolo 11, I-34131 Trieste, Italy
 \item Department of Astronomy, School of Science, SOKENDAI (The Graduate University for Advanced Studies), 2-21-1 Osawa, Mitaka, Tokyo 181-8588, Japan
 \item Laboratory for Multiwavelength Astrophysics, School of Physics and Astronomy, Rochester Institute of Technology, 84 Lomb Memorial Drive, Rochester, NY 14623, USA
 \item Dipartimento di Fisica e Scienze della Terra, Universit`a degli Studi di Ferrara, via Saragat 1, I-44122 Ferrara, Italy
 \item Instituto Nacional de Astrof\'isica, \'Optica y Electr\'onica, Tonantzintla, 72840 Puebla, Mexico
 \item Institute of Astronomy, University of Cambrige, Cambridge CB3 0HA, UK
 \item Facultad de Astronom\'ia y Geof\'isica, Universidad de La Plata, Paseo del Bosque s/n, B1900FWA La Plata, Argentina
 \item Department of Physics and Astronomy, University of California, Los Angeles, 430 Portola Plaza, Los Angeles, CA 90095, USA

\end{affiliations}

\clearpage
\noindent\textbf{\large Methods}
\vspace{-0.3cm}

\section{Cosmology and other definitions}
\vspace{-0.2cm}
Throughout this paper, we assume a flat $\Lambda$CDM cosmology with $\Omega_\mathrm{m} = 0.29$, $\Omega_\mathrm{\Lambda} = 0.71$ and $\rm H_0=69.6\,\rm km\,s^{-1}\,Mpc^{-1}$; and a Solar abundance of $12+\log\rm(O/H)=8.69$\cite{Asplund2009}.

\section{JWST/MIRI observations and data reduction}\vspace{-0.2cm}
Observations were conducted  as part of project GO-3703 (PI: J. Zavala) using the MIRI low resolution spectrometer (with the P750L filter) in slit mode. The target was observed in three different visits
using the FASTR1 readout pattern and 
121 groups per integration, 16 integration per exposure, and 1 exposure per specification with 2 dither positions ”along slit nod”. Each visit has an on-source time of 10828.26\,s (summing 9\,h in total). Target acquisition (TA) observations on a bright star (RA=00h13m58.3s; Dec=-30$^\circ$20$'$14.10$''$) were conducted before every visit to ensure the target is placed with subpixel accuracy ($<10\,$mas) at the nominal slit center location.

We reduce the MIRI data using the official \textit{JWST} pipeline (Version 1.13.4), Calibration Reference Data System (CRDS) version 11.16.16 and CRDS context \verb!jwst_1174.pmap! to assign the reference files. 
We adopt the stage 1 pipeline procedures unchanged, resulting in six count-rate images (2 dithers, 3 exposures each). Then, we run the \texttt{spec2pipeline} stage, which performs flux calibration and various instrument corrections, separately on the individual count-rate exposures, yielding six individual \texttt{s2d} images. Due to the presence of a bright nearby galaxy within the slit in one of the dither positions (see the top-left panel in Fig. 1), some residual emission is seen in half of the data. Therefore, we continue to treat each dither position separately, averaging the three exposures for each first and yielding one \texttt{s2d} file for each dither. Next, we perform a background subtraction on each dither, separately, rather than simply subtracting the two. This was done using the \texttt{Background2D} task in \texttt{photutils}, masking a $r=3.5$ pixel circular region around the detected emission lines. We estimate the median background using a box size of $2\times 2$ pixels. 
We then input these background-subtracted \texttt{s2d} files into the \texttt{spec3pipeline} stage, which performs outlier rejection via sigma clipping and combines the two dither positions into a single 2D spectrum. 
Finally, we perform 1D extraction manually using a boxcar filter with a width that scales with the MIRI PSF FWHM, as shown in the top panel of Fig. 1. Above $9\,$\um, where the noise r.m.s. per channel notably increases (due to the sensitivity of the detector and the higher spectral resolution), we re-bin the data with a 2-channel bin. 

An alternative reduction following the standard steps in the pipeline, results in a similar spectrum but with a continuum baseline slightly offset towards negative values. After correcting this systematic offset, the [OIII] line luminosity is around 20\% brighter than in our manual reduction, but with lower signal-to-noise ratio. Given that our modified reduction provides better r.m.s. noise across the whole spectrum, we adopt it for our analysis, but note that our results would not significantly change otherwise. 

\section{Spectroscopic redshift and line measurements}\vspace{-0.2cm}

\noindent{\bf Emission line fitting and line ratios:}
A single Gaussian function was used to fit the H$\alpha$ line assuming negligible contribution from the [NII] doublet. In the case of the [OIII] doublet, we perform a simultaneous two-Gaussian fitting. During this procedure we assume the same line-width  for the two lines (leaving it as a free parameter) and fix the  5007/4959 peak line ratio to 3. We use a Levenberg-Marquardt algorithm to perform  non-linear least-square minimization (using the {\sc lmfit} function).  The measured line-widths are consistent with the instrumental resolution\cite{Kendrew2015} of $R\equiv\Delta\lambda/\lambda\approx81$ around the position of the [OIII] doublet (vs $R\approx83$ from the measured line-width) and 132 around H$\alpha$ (with a slightly wider measured line-width corresponding to $R\approx103$).

To infer an upper limit on the H$\beta$ line flux density, we assume the 2.5$\sigma$ r.m.s value as the upper limit for the line peak and the same line-width as the [OIII]5007\AA\ line, which is determined by the line spread function of the instrument. This is a reasonable assumption since the lines are very close in wavelength space and thus the spectral resolution is expected to be very similar.  This results in a H$\beta$ line flux upper limit of  $\rm 2.0\times10^{-18}\,\rm\,erg\,s^{-1}\,cm^{-2}$.
Alternatively, the H$\beta$ line flux density was  inferred from the detected H$\alpha$ line. In the case of a dust-free environment and Case B recombination-line radiation (see details in ref.\cite{Osterbrock2006}), the expected H$\alpha$-to-H$\beta$ line ratio is 2.85 for $n_e=10^4\,\rm cm^{-3}$ and $T_e=10,000\,$K (note that this ratio does not change significantly for other values of density and temperature). Under these assumptions, we estimate an integrated line flux density of $\rm 0.9\times10^{-18}\,\rm\,erg\,s^{-1}\,cm^{-2}$.  
We adopt this value along the paper since we do not expect a significant dust attenuation for this galaxy (based on the SED fitting results, the dust continuum ALMA constraints\cite{Bakx2023}, and the blue UV slope\cite{Castellano2024}). Actually, adopting the Calzetti attenuation law\cite{2000ApJ...533..682C} and dust extinction values of $\rm A_V\approx0.1-0.3$ -- similar to the SED fitting results described below -- would decrease the estimated H$\beta$ line flux only by $\lesssim10\%$.

Finally, a $2.5\sigma$ upper limit for the [SII] doublet was derived using the local r.m.s value around $8.97\,\micron$ and adopting a line-width 5\% narrower than the H$\alpha$ line (following the expected spectral response of the instrument\cite{Kendrew2015}). All these measurements are reported in Table \ref{tab:simple-table}.

\noindent{\bf Spectroscopic redshift.}
The dominant uncertainty on the spectroscopic redshift comes  from the current MIRI/LRS wavelength calibration. While the current calibration accuracy is estimated to be around $\pm20\,$nm\footnote{(\url{https://www.stsci.edu/contents/news/jwst/2024/updates-to-miri-low-resolution-spectrometer-reference-files})}, this still can introduce a redshift offset of $\Delta z\sim0.04$. 
A recent calibration update (referenced as \verb!jwst_1174.pmap! in the JWST pipeline), introduced a shift of about 50\,nm at 6\,microns and decreasing to nearly zero at the red end of the spectral range. A similar correction but a bit less extreme in the blue part was independently found by ref.\cite{Beiler2023} using observations of a Y Dwarf. 
Fitting the line with a single Gaussian component, we obtain $z_{\rm H\alpha}=12.36\pm0.02$\,(random)$\pm0.04$\,(systematic) with the updated 2024 JWST calibration. For comparison, adopting the calibration from ref.\cite{Beiler2023}, we obtain $z_{\rm H\alpha}=12.37\pm0.02$\,(random)$\pm0.04$\,(systematic).  In the case of the [OIII] doublet, and using a two-Gaussian simultaneous fitting as described above, we obtain  $z_{\rm [OIII]}=12.29\pm0.01$\,(random)$\pm0.04$\,(systematic) using the most recent pipeline calibration, or $z_{\rm [OIII]}=12.33\pm0.01$\,(random)$\pm0.04$\,(systematic) using an  alternative wavelength calibration\cite{Beiler2023}. The differences between the inferred redshifts are mainly attributed to the current wavelength calibration uncertainty of $\sim20\,$nm. Across this paper we adopted the average of the redshifts obtained with the official pipeline calibration, resulting in $z=12.33\pm0.02$\,(random)$\pm0.04$\,(systematic) (or simply $z=12.33\pm0.04$, after taking the square root of the sum of the square of the two errors). Similar redshift constraints were derived from the NIRSpec observations\cite{Castellano2024} with a weighted redshift average between four different emission lines of $z=12.342\pm0.009$.
A more precise MIRI/LRS wavelength calibration in the future will allow to derive spectroscopic redshifts with better than 1\,\% precision even with the low resolution spectrometer.

\section{Inferred parameters from the emission lines.}

\begin{table*}
  \centering
  \begin{tabular}{|c|c|c|c|c|c|c|}
    \hline
    & H$\beta^\dag$ & [OIII]4959\AA &[OIII]5007\AA & H$\alpha$  & [SII]6717\AA,6731\AA & [OIII]88\um $^\ddag$ \\
    \hline
    Line flux [erg/s/cm$^2$]$\times10^{-18}$ & $<2.0$ OR $0.9\pm0.2$ & $1.6\pm0.2$  & $4.7\pm0.5$ & $2.5\pm0.7$ & $<2.6$ &$<0.25$\\
    \hline
  \end{tabular}
  \caption{Measured line flux densities or upper limits without correcting for the potential effect of gravitational amplification ($\mu\approx1.3$ according to ref.\cite{Bergamini2023}).  \\
  $^\dag$Two values are given: a $2.5\sigma$ upper limit directly constrained from the data, and the  inferred value from H$\alpha$ (valid in the case of zero dust attenuation; see text for details). \\
  $^\ddag$ Scaled from ref.\cite{Bakx2023} assuming a FWHM$=200\,$km/s.\\
  }
  \label{tab:simple-table}
\end{table*}

\noindent{\bf Metallicity:}
We adopted the recent calibration from ref.\cite{Sanders2024} based on a sample of 46 galaxies at $z\approx2-9$ observed with JWST/NIRspec and with multiple line detections (including temperature-sensitive lines), from which we derive 
$12+\log\rm(O/H)=7.40^{+0.52}_{-0.37}$ taking into account the uncertainty on the  $\rm [OIII]$-to-$H\beta$ line ratio and the observed scatter in the calibration sample. This corresponds to $Z=0.05^{+0.12}_{-0.03}\,Z_\odot$. 
Using instead the calibration from ref.\cite{Nakajima2022a}, calibrated using local analogs, results in a consistent metallicity of $\sim0.1\,Z_\odot$. We finally use the theoretical predictions for metallicity calibrations of ref.\cite{Hirschmann2023}, based on  the IllustrisTNG\cite{Pillepich2018} simulations connected to photo-ionisation models and focused only on galaxies at $z>4$. This theoretical calibration implies a metallicity of $Z=0.04\pm0.02\,Z_\odot$). 

Note that our assumption of a negligible dust attenuation when estimating the H$\beta$ line luminosity does not bias the inferred metallicity. This is due to the turnover of the relation at higher metallicities where lower values of $\rm [OIII]$-to-$H\beta$ are expected (while dust attenuation would imply higher line ratios). This turnover implies, however, a second solution for GHZ2 of  $Z\approx0.55\,Z_\odot$. This value is inconsistent with the independent constrains obtained from the NIRSpec data\cite{Castellano2024}, with several diagnostic suggesting values around or below $0.1\,Z_\odot$. We can thus conclude that the  gas-phase metallicity in GHZ2 is close to the aforementioned value of $Z=0.05^{+0.12}_{-0.03}\,Z_\odot$; relatively low compared to lower redshift galaxies, but not pristine despite its young age.

\noindent{\bf Balmer decrement:} In the presence of dust, the $H\alpha$-to-H$\beta$ line ratio is expected to deviate from the theoretical value since the short-wavelength H$\beta$ transition is more susceptible to dust attenuation. However, while the upper limit on the H$\beta$ line luminosity is consistent with a dust-free scenario it does not rule out the possibility of dust attenuation.  Deeper observations will be necessary to put any constraints on the presence (or absence) of dust in this galaxy via the Balmer decrement.

\noindent{\bf SFR and ionizing photon production efficiency.} As mentioned in the main text, we estimate a SFR of $12\pm4$\,M$_\odot$\,yr$^{-1}$ (or $9\pm3$\,M$_\odot$~yr$^{-1}$ after taking into account the gravitational magnification of $\mu=1.3$ estimated by ref.\cite{Bergamini2023}) using the calibration\cite{Reddy2022} $\rm SFR/L(H\alpha) = 10^{-41.67} (M_\odot\,yr^{-1})/(erg\,s^{-1})$. This calibration was derived from $Z=0.001$ BPASS population synthesis models with an upper-mass IMF cut of $100\,M_\odot$ and including the effects of stellar binaries. Using instead the relation of ref.\cite{Kennicutt2012}, calibrated for lower redshift systems with close-to-solar metallicity, results in a higher SFR by a factor of $\sim2.5$. This difference is mainly attributed to the absence of low-metallicity stars  and binary star interactions that produce higher ionization photons.  
In addition, we infer the SFR from the SED modelling as described below.

\begin{figure*}[hbt!]
  \begin{center}\hspace{-0.2cm}
  \includegraphics[width=0.52\linewidth]{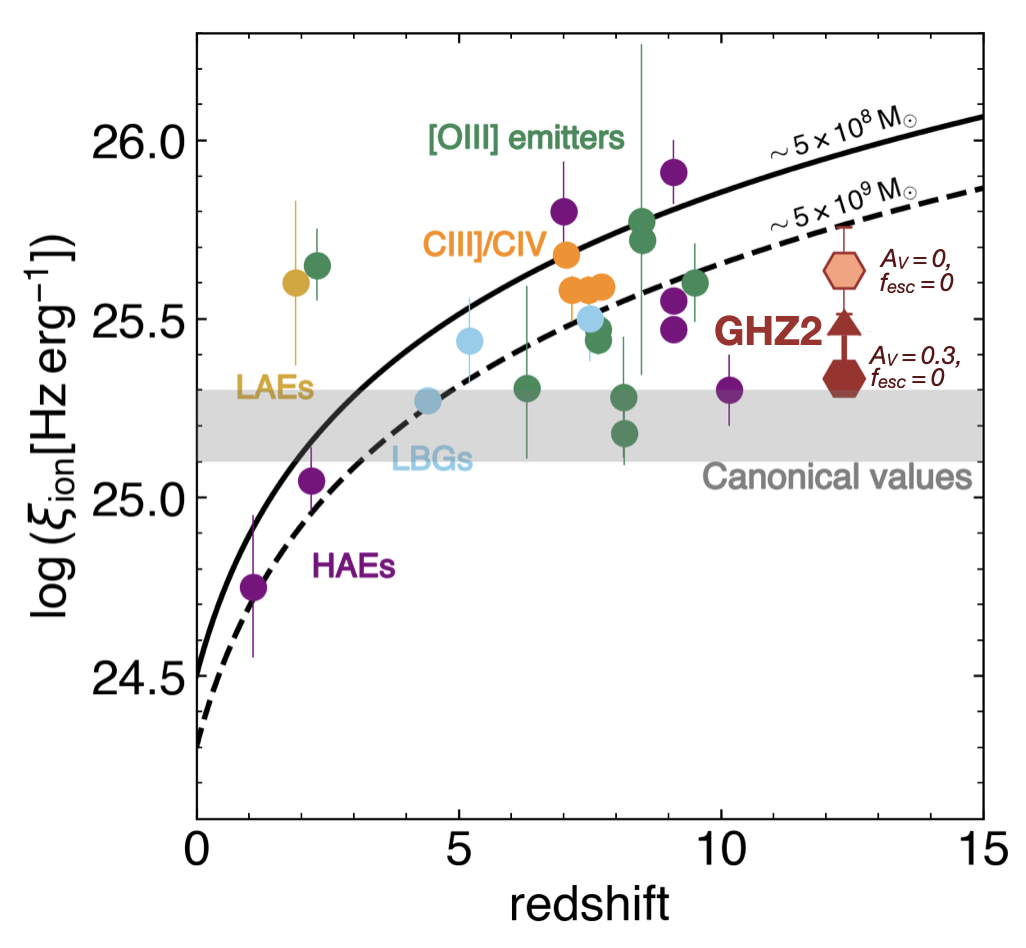}\vspace{-0.7cm}
  \end{center}
    \caption{\small \textbf{Ionizing photon production efficiency.} Inferred ionizing photon production efficiency for GHZ2 assuming a conservative dust attenuation  of $\rm A(V)=0.3\,$mag (illustrated with the solid hexagon). For comparison, we show the inferred value for a zero dust attenuation (light red) and estimated values for other samples (average values\cite{Matthee2023} and individual measurements\cite{Rinaldi24,Alvarezmarquez23,Atek23,Fujimoto23,Hsiao24,Lin23,Stark15,Stark17}), along with the redshift evolution from ref.\cite{Matthee2017}. Error bars represent $1\sigma$ errors in the case of individual measurements or the scatter of the measurements for samples' average. The bright H$\alpha$ emission of GHZ2 implies a high ionizing photon production, likely above the typical values adopted for galaxies contributing to the reionization process\cite{Robertson2013}.  }
    \label{fig:ion_efficiency}
\end{figure*}

The H$\alpha$ line luminosity was also used to estimate the ionizing photon production efficiency, $\xi_{\rm ion}$, following ref.\cite{Matthee2017}, which is related to the number of produced ionizing photons per UV luminosity (or SFR). To be conservative, during this calculation we assume an escape fraction of $f_{\rm esc}=0$ (any other value above zero will result in a higher  $\xi_{\rm ion}$) and a dust attenuation of $\rm A(V)=0.3\,$mag (note that the attenuation of the young stellar population inferred from our fiducial SED modelling is $\rm A(V)=0.1\,$mag, with other models suggesting even lower values). Under these assumptions, we  estimate an ionizing photon production efficiency of GHZ2 to be $\xi_{\rm ion}\gtrsim2\times10^{25}\rm\,Hz\,erg^{-1}$, as shown in Extended Data Fig. \ref{fig:ion_efficiency}.

\section{Gravitational lensing modelling}
A detailed strong-lensing model of the galaxy cluster Abell 2744 predicted\cite{Bergamini2023} a gravitational amplification of $\mu=1.3$ for GHZ2/GLASS-z12. Nevertheless, the presence of other bright galaxies along the line-of-sight (between the cluster and the target) might produce second-order effects. To test this, we modeled the total mass distribution of the closest galaxies as foreground lenses with one-component models, namely  Singular Isothermal Sphere (SIS) profiles. Their redshifts and stellar masses were extracted from the GLASS catalog\cite{Paris2023}, spanning $z\sim1.5-3.5$ and $M_\star= 8\times10^7 - 2\times10^9\,M_\odot$. The closest source is the most massive galaxy, with a spectroscopic redshift of  $z=1.682$. We assume a conservative effective velocity dispersion value of $100\,\rm km/s$ for all the galaxies, corresponding to the highest expected value given their stellar masses\cite{Grillo2014}. This was then used to estimate the associated Einstein angle for a source at the redshift of GHZ2 ($z=12.33$). Finally, to asses the potential effect of this secondary gravitational lensing amplification, this angle was compared with the angular separation
between the corresponding perturber mass center (after correcting its position for the deflection of the lens cluster) and the position of GHZ2/GLASS-z12 on the perturber's plane.

Based on the this analysis, we found that the angular separation between GHZ2 and the closest galaxy on its plane exceeds by more than five times ($\theta/\theta_{\rm E}\approx5.4$) the estimated Einstein angle (being 8-20x larger for the other foreground galaxies). While a simple SIS gravitational lens model would imply a magnification of $(1-1/5.4)^{-1}\approx 1.23$ from this close galaxy, the total magnification experienced by GHZ2 cannot simply be obtained by multiplying the magnification factor of the cluster by that of the galaxy (which would results in $\mu\approx1.6$). This can only be accurately quantified using a multiplane lensing approach which is beyond the scope of this paper. Therefore, to avoid introducing biases relative to this modelling, we decided to adopt the value of\cite{Bergamini2023} $\mu=1.3$, but notice that the uncertainties on this value allows for slightly larger magnification values with increments in the order of $\sim20-30\%$.

\section{SED fitting}
The broad-band spectral energy distribution, jointly with the spectroscopic measurements of the H$\alpha$ and [OIII] emission lines, were fitted to stellar population and nebular gas emission models to estimate the stellar mass, SFR, and mass-weighted age of GHZ2. The NIRCam photometry is slightly different from previous estimations\cite{Castellano2022} since we include new observations obtained  in July 2023 (see ref.\cite{Castellano2024} and Merlin et al. in prep. for further details). We used the {\sc synthesizer-AGN}  code \cite{2003MNRAS.338..508P,2008ApJ...675..234P,2024arXiv240108782P}, with the stellar populations models from ref.\cite{2003MNRAS.344.1000B} and with a Chabrier\cite{2003PASP..115..763C} initial mass function with stellar mass limits between 0.1 and 100~$\mathrm{M}_\odot$. We probed all sub-solar metallicity models. The star formation history was set to a double burst, each stellar population described by a delayed-exponential law with possible timescales ranging from 1 to 100~Myr, ages from 0.1~Myr to the age of the Universe at $z=12.36$. Each stellar population is allowed to be affected by independent dust attenuations described by Calzetti law\cite{2000ApJ...533..682C}, with A$_{\rm V}$ values ranging from 0 to 1~mag. Nebular continuum and line emission was modeled with {\sc Cloudy} version c23.0.1 \cite{1998PASP..110..761F,2023RMxAA..59..327C}, assuming 10,000~K gas with $10^4$~cm$^{-3}$ density and abundances linked to the stellar metallicity and ionizing photon flux provided by the stellar models. The main derived properties  (without correcting for the gravitational amplification) are: stellar mass $\log\mathrm{M_\star/M}_\odot=9.03^{+0.13}_{-0.28}$, SFR$_\mathrm{10Myr}=7\pm2$~M$_\odot$~yr$^{-1}$, mass-weighted age $28^{+10}_{-14}~$Myr. The models also support the  high ionization parameter inferred from the BPT diagram, with a best-fit value of $\log \mathrm{U}=-1.1\pm0.4$, and a consistent stellar metallicity of $\mathrm{Z/Z_\odot}=0.020^{+0.030}_{-0.015}$. The attenuation for older and younger star populations is constrained to be $\mathrm{A(V)}= 0.3^{+0.1}_{-0.2}~$mag and $0.1^{+0.2}_{-0.1}~$mag, respectively. The best-fit SED obtained from this analysis is presented in the Supplementary information.

Additionally,  we use the {\sc bagpipes}\cite{Carnall2018}  and  {\sc cigale}\cite{Boquien2019} SED-fitting codes with the same stellar population models, IMF, and dust attenuation law. In the case of {\sc bagpipes} we adopt a "bursty continuity" model for the star formation history and found a  general good agreement in the age of the stellar population ($30^{+40}_{-20}~$Myr) with a very similar fraction of mass formed during the last 30$\,$Myr. Other parameters are also consistent within the error bars ($\log \mathrm{U}=-1.4^{+0.2}_{-0.3}$;  SFR=$2^{+3}_{-1}$~M$_\odot$~yr$^{-1}$; $\log\mathrm{M_\star/M}_\odot=8.38^{+0.23}_{-0.18}$), with the major difference being the close-to-zero attenuation ($\mathrm{A(V)}= 0.01^{+0.02}_{-0.01}~$mag) and a slightly higher metallicity ($Z=0.22^{+0.06}_{-0.05}\,Z_\odot$).
Similarly, the results with {\sc cigale} implies a mass-weighted stellar age of $26\pm55\,$Myr, $\log\mathrm{M_\star/M}_\odot=8.3\pm0.3$,  $\log \mathrm{U}=-2.1\pm0.5$, $\mathrm{A(V)}= 0.11^{+0.13}_{-0.11}~$mag, and SFR$_\mathrm{10Myr}=15\pm9$~M$_\odot$~yr$^{-1}$, with an even higher (but largely uncertain) metallicity of $Z=0.56\pm0.20\,Z_\odot$. Note that all these values are not corrected for gravitational amplification.  

\section{Photoionization and line emissivity models} The photoionization models presented in Fig. \ref{fig:MZR}, showing the ratio between [OIII]5007\AA\ and H$\beta$ and between [NII]6583\AA\ and H$\alpha$, are taken from ref.\cite{Nakajima2022b}.  They were calculated using the spectral synthesis code, {\sc cloudy}\cite{Ferland2017}, and assuming the stellar population models from BPASS, including binaries systems, with a Kroupa\cite{} IMF with an upper mass cut of $300\,\rm M_\odot$ (although similar results are obtained for $M_{\rm up}=100\,\rm M_\odot$). In these models, the ionization parameter, defined as the ratio between the hydrogen-ionizing photons and the number density of hydrogen atoms, was varied from -3.5 to -1.0, while the gas electron density was fixed to $10^3\rm\,cm^{-3}$.  As seen in Fig. \ref{fig:MZR}, the  [OIII]5007\AA-to-H$\beta$ line ratio of GHZ2 is reproduced only by models with high ionization parameter above -2.0 (in the case of $Z\approx0.1-0.2\,Z_\odot)$. At lower metallicities ($Z\lesssim0.05\,Z_\odot$), the models predictions lie below the inferred value and would require  harder ionizing radiation to explain the observational constraints.
For instance, AGN-driven models show higher [OIII]5007\AA-to-H$\beta$ ratios (by $\sim0.3\,$dex), when compared with the stellar-driven models at fixed metallicity\cite{Nakajima2022b}.

On the other hand, the predicted line ratios between the two transitions of the double-ionized oxygen ([OIII]5007\AA\ and [OIII]$88\micron$) shown in Fig. \ref{fig:ALMA_constraints} were generated 
using the PyNeb getEmissivity package\cite{Luridiana2015},  with the default atomic data. The ratio was calculated for different values of electron density and temperature ranging from $\log(n_e\rm[cm^{-3}])=0-10$ and $\log(T_e\rm[K])=3.7-4.5$. As can be seen in the figure, the ratio is highly sensitive to electron density with a milder dependency on electron temperature (given the significant different energy levels of the two transitions), but independent of metallicity since both transitions arise from the same ion.

\section{Revisiting the ALMA data and constraining the electron density}
\begin{figure*}[hbt!]
  \begin{center}\vspace{0.5cm}
  \includegraphics[width=0.5\linewidth]{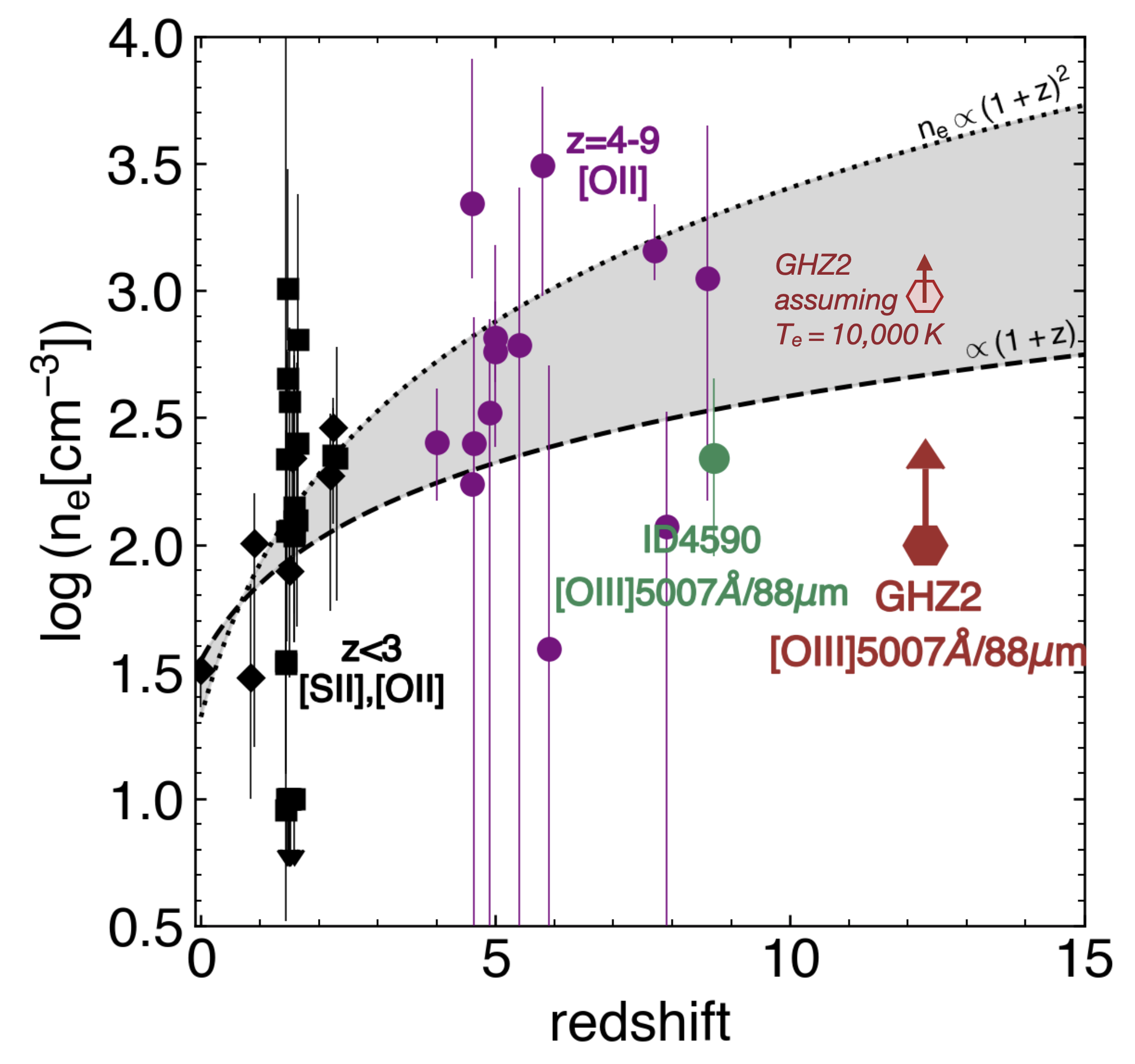}\vspace{-0.6cm}
  \end{center}
    \caption{\small \textbf{The redshift evolution of the electron density}. Compilation of spectroscopically-derived measurements (with $1\sigma$ error bars) of ionized gas electron density in galaxies at different redshifts (modified from ref.\cite{Isobe2023} and including other measurements\cite{Kaasinen2017,Fujimoto2024}) along with the constraints inferred for our target, GHZ2, at $z=12.33$. Our results support the evolution towards higher electron densities at high redshifts, which might be associated with the high ionization parameters and high star formation rate surface densities of the bright population of high-redshift galaxies recently discovered with JWST. 
    }
    \label{fig:ne_redshift}
\end{figure*}

The initial investigation of GHZ2/GLASS-z12 at sub-mm wavelengths with ALMA only revealed a tentative line\cite{Bakx2023} at $258.7\,\rm GHz$, claimed to be [OIII]88$\mu$m at $z=12.117$. 
This is inconsistent with the MIRI and NIRSpec spectroscopic redshift, implying that the line was not real or, at least, not associated with GHZ2. With the updated redshift information, we visually inspect the 254.35\,GHz region  and fit 
the extracted spectrum using a single Gaussian function, centered at the peak channel. The best-fit implies a line flux of $\sim2.5\times10^{-19}\rm\,erg/s/cm^{-2}$ with $\rm FWHM\approx200\,\rm km\,s^{-1}$, which is consistent with the previously reported\cite{Bakx2023,Popping2023} upper limit, after taking into account the different line-widths (originally adopted to be $\approx100\,\rm km\,s^{-1}$ in the previous ALMA analysis\cite{Bakx2023}). Given  the low statistical significance ($\sim3\sigma$) of this measurement, to be conservative, we treat it here only as an upper limit. 
This implies a [OIII]5007\AA-to-[OIII]88$\mu$m  line ratio above $19$ (or $\log(\rm[OIII]_{5007A}/[OIII]_{88\mu m})\gtrsim1.3$) and  is used in Fig. \ref{fig:ALMA_constraints} to obtain a  conservative lower limit on the electron density of $n_e>100\,\rm cm^{-3}$, which varies as a function of electron temperature. Assuming the typical value of  $T_e=10,000\,$K, the current constraints imply an extreme $n_e$ value above $1,000\,\rm cm^{-3}$. This is consistent with the picture of higher electron densities at high redshifts, as shown in Extended Data Fig. \ref{fig:ne_redshift}.

High electron densities would naturally explain other non-detections of the [OIII]88$\mu$m line\cite{yoon2023} at $z>10$ given its relatively low critical density of\cite{Osterbrock2006} $n_c\approx5\times10^2\,\rm cm^{-3}$. Above this value, collisional deexcitation plays a significant role in gas cooling, diminishing the line luminosity produced by radiative cooling that is more efficient in the low density regime. Hence, future ALMA follow-up may be more successful by targeting the [OIII]52$\mu$m line thanks to its higher critical density of $n_c\approx3\times10^3\,\rm cm^{-3}$.\\



 

\clearpage
\noindent{\bf \large References}\vspace{10pt}
\bibliography{bibMake}

\clearpage
\onecolumn
\noindent\textbf{\large Extended Data Figures} 

\clearpage
\onecolumn
\pagenumbering{gobble} 
\noindent\textbf{\large Supplementary Information} 
\vspace{1cm}

\begin{figure*}[h]
  \begin{center}
  \includegraphics[width=0.8\linewidth]{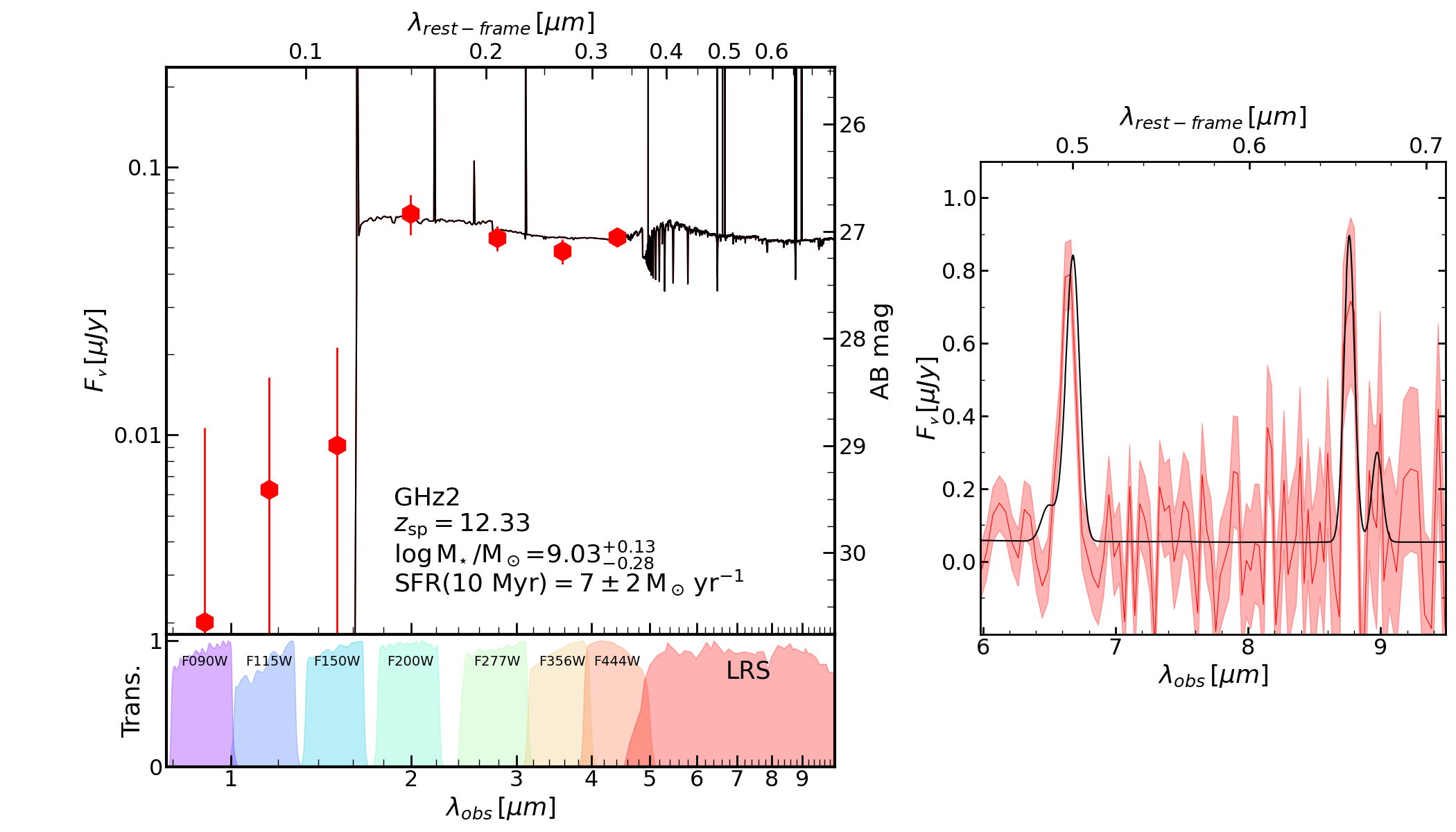}
  \end{center}
    \caption{\small \textbf{The best-fit spectral Energy Distribution of GHZ2/GLASS-z12}. Example of our photo-spectroscopic analysis combining the NIRCam photometry (illustrated by the red points on the left panel with associated $1\sigma$ uncertainties) and the MIRI/LRS spectra (illustrated by the red solid line on the right panel and shaded region showing $1\sigma$ uncertainties). The best-fit SED obtained with the {\sc synthesizer} code is shown by the black solid line (original spectral resolution on the left panel, convolved to the resolution of MIRI/LRS on the right panel) along with some of the best-fit parameters (without correcting for gravitational amplification). Note that  {\sc cigale} and  {\sc synthesizer} only use the line fluxes information in the fitting (along with the measured photometry), while {\sc bagpipes} uses the full spectrum as an input. }
    \label{fig:sed}\vspace{-0.8cm}
\end{figure*}

\end{document}